\documentstyle[aps,epsf]{revtex}

\tightenlines

\begin{document} \renewcommand{\thefootnote}{\fnsymbol{footnote}}

\draft \twocolumn[
\hsize\textwidth\columnwidth\hsize\csname@twocolumnfalse\endcsname \title{\bf
Magnetic properties and Hall effect of $Pr_{0.7}Sr_{0.3}MnO_3$ \\single crystal}

\author{D. Dimitrov\footnote{Email: dimitrov@issp.bas.bg}$^1$,  V.
Lovchinov\footnote{Email: lovcinov@issp.bas.bg}$^{,1}$ M. Gospodionov$^1$ \\Ph.
Vanderbemden\footnote{Email: Philippe.Vanderbemden@ulg.ac.be}$^{,2}$  \\and \\M.
Ausloos\footnote{Email: Marcel.Ausloos@ulg.ac.be}$^{,3}$}

\address{$^1$  Institute of Solid State Physics, Bulgarian Academy of Sciences,
1784 Sofia, Bulgaria \\ $^2$  S.U.P.R.A.T.E.C.S., Universit\'e de Li\`ege,
Institut d'Electricit\'e Montefiore,  B28  ÊSart Tilman, B-4000 Li\`ege, Belgium
\\ $^3$ S.U.P.R.A.T.E.C.S., Universit\'e de Li\`ege, Institut de Physique,  B5
Sart Tilman, B-4000 Li\`ege, Belgium}

\date{\today}


\draft

\maketitle

\begin{abstract} {Measurements of  DC  magnetization, AC susceptibility and Hall
effect of $Pr_{0.7}Sr_{0.3}MnO_3$ single crystal are reported. The various
features are discussed in terms of competing electronic and spin ordering phase
transitions, in particular concerning the $e_g$ and $t_{2g}$ competition and its
effect on spin ordering.}

\end{abstract} \vskip 0.5cm ]


\newpage

\noindent

{\bf I. Introduction} \vskip 12pt

Renewed interest in perovskite manganites is due to the very rich physical
phenomena, e.g. colossal magnetoresistance, charge ordering and orbital ordering
which they exhibit\cite{review}.   The resulting rich phase diagrams derive from
strong correlations between spin, charge, and lattice degrees of freedom. As a
function of temperature    different magnetic and electrical phases are found,
like paramagnetic insulating, ferromagnetic metallic, antiferromagnetic
insulating, canted antiferromagnetic metallic, and weak ferromagnetic insulating.
The extension of these regions  also depends on the value of the applied magnetic
field. Electronic properties, especially near the insulator-metal (IM) phase
boundary, are sensitive to the magnitude of the single-electron bandwidth ($W$).
In transition-metal oxides with a distorted perovskite structure ABO$_3$, where A
and B are the rare-earth and transition-metal ions, respectively, the $W$ value
can be modified by varying the Mn-O-Mn bond angle\cite{2}  through change in the
average  A and B ionic radius. A criterion for tilting of the MnO$_6$ octahedra
is the tolerance factor $t=(r_A+r_O)/ (r_B+r_O)$, where $r_A$ and $r_B$ are the
radii of A- and B-site cations and $r_O$ is the oxygen radius. A small $t$ value
results in a reduction of the Mn-O-Mn bond angles, whence a narrowing of the
$e_g$ bandwidth, and {\it in fine} localization of the charge carriers. In
general, perovskite-type manganites A$_{1-x}$B$_x$MnO$_3$ become conducting
ferromagnets upon hole doping ($x$).

The main mechanism controlling the magnetic and electronic properties is the so
called double exchange (DE), a mechanism in which the magnetic coupling between
neighboring Mn$^{3+}$ and Mn$^{4+}$  ions is mediated through the transfer of an
electron with spin memory. These itinerant $e_g$ carriers are magnetically
coupled with the local  $t_{2g}$ spins via the $single-site$ exchange interaction
through HundÕs rule coupling $J_H$.

In this sense, the giant magnetoresistance (GMR) which the doped manganites show
around the Curie temperature ($T_C$) is ascribed to the alignment of the local
$t_{2g} $ spins and to the reduction of the spin scattering for the $e_{g}$
carriers by the magnetic field\cite{4}. However, the crystal lattice also plays a
role in this phenomenon through the Jahn-Teller distortion associated with
Mn$^{3+}$  ion, - a distortion which is $static$ in the orbitally ordered end
member LaMnO$_3$ but becomes $dynamic$\cite{5} in the ferromagnetic composition
range, e.g. when La is replaced by a divalent atom like Ca or Sr.

In addition to the ferromagnetic transition, the doped manganites often undergo a
phase transition to a charge-ordered (CO) state, in which the nominally Mn$^{3+}$
and Mn$^{4+}$  species show a real-space ordering in the crystal, e.g. in the
case where the doping level ($x$) is near a commensurate value
(e.g.=1/2)\cite{6,7,8}. The charge-ordering (CO) transition is accompanied by an
appreciable change in the lattice parameters\cite{9,10}.  These CO transitions
are classified\cite{4} into two types. The first type is described as a
conducting ferromagnetic system which becomes an antiferromagnetic CO insulator
when the temperature decreases. For the second type, the CO state emerges at
higher  temperatures where the ferromagnetic metallic state is no longer present.

One of the most interesting compounds which shows quite surprising
properties\cite{3} is Pr$_{0.7}$Ca$_{0.3}$MnO$_3$, for which $r_{Pr}$=0.1 nm and
$r_{Ca}$=0.104 nm, respectively. If calcium is fully replaced by
strontium\cite{lin,171,qili,075a,075b,104,films067} ($r_{Sr}$=0.12 nm) the
tolerance factor $t$ should be reduced by a few \%, but tiny changes in such
materials drastically modify properties\cite{lin}, whence physical properties of
Pr$_{0.7}$Sr$_{0.3}$MnO$_3$ should be interesting.  Following the above
classification we will find out that  our Pr$_{0.7}$Sr$_{0.3}$MnO$_3$ samples are
of the  first type CO transition. However before reporting investigations on
Pr$_{0.7}$Sr$_{0.3}$MnO$_3$, let us recall basic considerations in this research
field.

Numerous investigations trying to unveil the nature of the charge transport have
been published in the recent years, but the scattering mechanisms and their
relative strength or importance are still debated upon\cite{review}. One of the
main accents of discussion is placed on the correlation between the magnetic
order and the charge scattering, which can be on a single spin or on a collective
excitation, i.e. a magnon or a set of magnons\cite{Hubert}.  In this regard, on
one hand, the transverse magnetoresistance, i.e., the Hall effect, - and the
electrical resistivity, should provide some valuable information, like on the
relative importance of the number and mobility of charge carriers, hint to
specific scattering mechanisms, etc. On the other one should consider the
magnetic susceptibility, - and the magnetization. Thus concomitant considerations
of both magnetic measurements and Hall effect studies are the aims of this paper
to elucidate the physical picture occurring in the $Ca$-free
Pr$_{0.7}$Sr$_{0.3}$MnO$_3$ system.

\vskip 12pt \noindent {\bf II. Experimental}

\vskip 12pt

The single crystals were produced by the HTSG (High Temperature Solution Growth)
method. The synthesis of the main substance Pr$_{0.7}$Sr$_{0.3}$MnO$_3$ was
carried out in a pure platinum crucible at 1280 C for 72 h. As a result, the
single crystals have a parallelepipedic form with maximal dimensions  3x3x2 mm
and density $d$=5.6 g/cm$^3$. Standard EDAX analysis was carried out and the
component contents were determined to be close to the expected value within the
EDAX usual error bars, i.e. $\pm$ 5\%. A monophase sample with perovskite
structure was acquired as seen through results of a  powder x-ray-diffraction.
The crystal structure was found to be pseudocubic (P$_{n3m}$) with a lattice
parameter $a$=0.7696 nm. More information about the sample preparation and
characterization is published elsewhere\cite{11}.

Transport phenomena have been measured using a standard four probe Van der Pauw
technique\cite{12} which allows measuring specific resistivity and Hall effect of
samples of arbitrary shape. A constant magnetic field of 0.7 T (which can be
switched in both directions) was utilized. This was the maximal value which our
experimental set-up allows. The temperature was measured by means of
copper-constantan thermocouples. The data accuracy is restricted to 2\% for the
electrical conductivity and 15\% for the Hall measurements, according to standard
error bar estimates. Below 270 K the electrical resistivity (not shown)
decreases, - {\it a posteriori} humps can be observed as hinted from the
subsequent magnetic data as seen in figures below. Above 270 K a transition  to a
metallic type conductor is well observed through a $quasi$ linear dependence of
the resistivity as a function of temperature; the position of the transition is
poorly defined in such a data though it can be observed that it is slightly field
dependent.

All magnetic measurements were performed in a Quantum Design PPMS when it worked,
following previously described routines\cite{VDB}. Before each measuring
sequence, the remnant field of the superconducting magnet was eliminated by the
standard practice of applying a succession of decreasing fields in alternate
directions. The saturation magnetization ($M_s$) has been investigated at liquid
Nitrogen temperature (77.3 K). Results presented on Fig. 1 show that the
saturation is achieved at fields higher than 1 T. Magnetization of 0.7 T (field
of our Hall effect studies) differs from $M_s$ with less than 3\% which makes the
reported data below adequate enough  for a meaningful discussion.

The DC magnetization ($M$) data were taken in the temperature range 10-300 K (FC
and ZFC) at 1 T magnetic field. AC susceptibility, $M^{'}$ and $M^{''}$, i.e. the
in-phase and the out of phase components, respectively were measured without, as
well as in presence of,  an external magnetic field with induction of 1 T from
300 to 10 K with step of 1 K in rate of 1 K/min; the AC field has a 1 mT
amplitude while $f$ = 1053 Hz. The applied magnetic field supplies important
information and helps to visualize different effects, e.g. CO modulation as a
function of temperature.

\vskip 12pt \noindent

{\bf III. Results and discussion} \vskip 12pt

\subsection{Results}

A typical magnetization $M(T)$  measured for a   1T DC magnetic induction  $vs.$
temperature is shown in Fig. 2. The curve at temperatures higher than 180 K
denotes  some peculiarities in the magnetic characteristics of the sample.  The
$M(T)$ curve (Fig. 2) presents an inflection in the region near 220 K.
Extrapolation of the $M(T)$ data to zero  allows to define the Curie temperature
$T_C$, where the transition between ferromagnetic and paramagnetic states takes
place,  to be about 270 K.

Hall effect investigation results  are presented on Fig. 3 for the temperature
range 77 Ð 380 K in the case of a magnetic field with a 0.7 T induction. With
increasing the temperature up to 170 K the Hall coefficient value remains nearly
zero. In the region 170-220 K the curve shows a sudden negative decrease followed
by a rapid increase up to 270 K.  The data falls back to about zero thereafter.

The magnetic susceptibility taken for an external field corresponding to a 1 T
induction  is plotted versus temperature on Fig. 4.  Two peaks are noticed after
a smooth decay, i.e. near  150 K and 270 K respectively.  Local minima are found
near 225 K and 150 K.

\subsection{Discussion}

From the susceptibility and the magnetization, it is clear that the system
reaches an ideal ferromagnetic ordering near 270 K in agreement with HundÕs rule.

All variations of the Hall coefficient can be explained following the reasons
expressed by Moritomo\cite{4} et al. for  Pr$_{0.7}$Ca$_{0.3}$MnO$_3$. The same
behavior can be observed for the Pr$_{0.7}$Sr$_{0.3}$MnO$_3$ system. At low
temperatures the interaction energy, after HundÕs rule, is quite low to start the
generation of itinerant $e_{g}$  carriers, i.e. there are no valence electrons
with great S and L quantum numbers. In this region the sample is an insulator. At
temperatures higher than 170 K $e_{g}$  carriers with great S and L numbers arise
and the system starts to order following HundÕs rule. Generation of carriers
begins and the compound converts into a semiconductor state. At the end of this
transformation the sample is antiferromagnetic. Further tilting of the MnO$_6$
bonds at angles and lengths prolongs with a temperature increase. Thus, after a
region of dynamic equilibrium between generation and recombination of carriers
the system undergoes a  semiconductor Ð metal phase transition which is a CO
transition for this compound.

The  reasons for the above as described interaction mechanisms can be confirmed
from the magnetic measurements, e.g. the AC susceptibility (Fig.4). As a function
of decreasing temperature, two large peaks, at 270 K and 200 K are indeed noticed
together with a rising contribution starting from zero near 170 K. One might
consider that  each peak width might be due to inhomogeneities and defects in the
sample, but this is unlikely in view of the characterization studies, or to an
extended range for fluctuations and intrinsic spin ordering phase competitions.

This question and mechanisms can be elucidated from the analysis of measurements
conducted under an external magnetic field. It should be expected that such 	a
field has significant effects not only on the ferromagnetic transition but on the
charge-ordering transition as well. This is because the effective transfer
integral ($t$) of the $e_{g}$  carriers strongly depends on the alignment of the
$t_{2g}$ spins\cite{15} $t=t_0 cos(\phi/2)$, where $t_0$ is the bare transfer
integral without spin scattering and $\phi$ is the relative angle of the
neighboring $t_{2g} $ spins. An external magnetic field would forcefully align
the $t_{2g}$ spins (such that $\phi$=0), whence enhances the carrier mobility,
$and$ modifies the CO state. In an extreme case, the application of a magnetic
field would destroy the CO state and induces  a FM state from the low-temperature
side, as exemplified\cite{10,16} in Pr$_{1-x}$Ca$_x$MnO$_3$ in fact.

Moreover the maximum at about 270 K indicates a fully completed ferromagnetic
state and determines the Curie temperature for the sample, i.e. 270 K. Further
increase in temperature initiates the transition toward the paramagnetic phase.
Carrier generation occurs above 270 K because of spin decoupling from the
external field. This mechanism is accompanied by a degradation of the magnetic
coupling between $e_{g}$  carriers thereby  favoring a  transition to a
semiconductor state. This effect is ''compensated'' by an interaction between
pair of electrons with $t_{2g} $ spins. A temperature increase intensifies  the
process of  spin coupling degradation leading to a   decrease of the compensating
effect.  Notice that on the lower side of the 270 K peak, the  spin  parallelism
is not necessarily perfect either, with large fluctuations, appearing like local
somewhat stable spin cluster inhomogeneities, as illustrated by the  occurrence
of a local minimum near 225 K. At 225 K a FM arrangement is reached which is
actually a semiconductor to metal transition.

\vskip 12pt \noindent

{\bf IV. Conclusions} \vskip 12pt

One of the most important problems in the physics of manganites is the
determination of the charge carrier density. The divalent atoms act as acceptors
so that the holes play the role of the charge carriers here\cite{17}. The
variation in  charge carriers and Hall density determine different mechanisms
(such as CO, AFM insulating states, FM or `cluster glass` state) connected with
magnetic, electrical and structure properties. Thus, some knowledge of the
competition between these factors may give some better understanding of the CMR
phenomenon. In the present investigation we have found that the effective ÒHall
densityÓ $n_H=1/(eR_0)$, including its sign, varies with temperature (Table 1).

Such strong variations are connected with the appearance of itinerant carriers.
Changes in their localization  strength or type  take place simultaneously with
magnetic rearrangements during which the charge can even be changed from hole to
electron. From such studies it appears that the carrier mobility is not the most
relevant parameter. We should like to emphasize the good agreement between the
Hall effect and the magnetic results. Especially, it seems that the
susceptibility in an external field gives a clear picture and serves to elucidate
the complicated interaction of the $e_{g}$  electrons and $t_{2g} $ spins,
inherent to  the perovskite manganites. This confirms and precises work on
polycrystalline large grain samples\cite{VDB}.

\vskip 0.6cm

{\bf Acknowledgments}

\vskip 0.6cm

We would like to thank Prof. Dr. M. Mihov from the Physics Department of  Sofia
University, and K. Zalamova for useful discussions  and support. A bilateral
grant from CGRI and BAS has allowed some exchange visit between Sofia and Li\`ege
groups which appreciably influenced the paper.

\newpage \noindent {\bf Figure captions}

\vskip 1 truecm \vskip 12pt \noindent {\bf Fig.1} DC magnetization for a
Pr$_{0.7}$Sr$_{0.3}$MnO$_3$ single crystal at liquid nitrogen temperature (77.3
K) as a function of the DC applied magnetic field

\vskip 1cm   \noindent {\bf Fig.2} Pr$_{0.7}$Sr$_{0.3}$MnO$_3$  sample magnetic
moment in a 1 T DC field   versus temperature

\vskip 1cm  \noindent {\bf Fig.3} Hall effect investigation of a
Pr$_{0.7}$Sr$_{0.3}$MnO$_3$ single crystal in a 0.7 T magnetic field

\vskip 1cm   \noindent {\bf Fig.4} AC magnetic susceptibility from the in phase
AC magnetic moment,   in a 1 T DC external induction

\vskip 1cm \noindent {\bf Table caption}

\vskip 12pt \noindent Table I:  Effective Hall density at various temperatures
for a Pr$_{0.7}$Sr$_{0.3}$MnO$_3$ single crystal

\vskip 1cm  


\begin{centering}

\begin{tabular}{|c|c|} \hline Temperature & Effective Hall density \\ (K) & $
(10^{20} cm^{-3})$  \\ \hline 77-150 & 4 \\ 150 & 28 \\ 170 & -24 \\ $>250$ & 1
\\ \hline \end{tabular}

\end{centering}


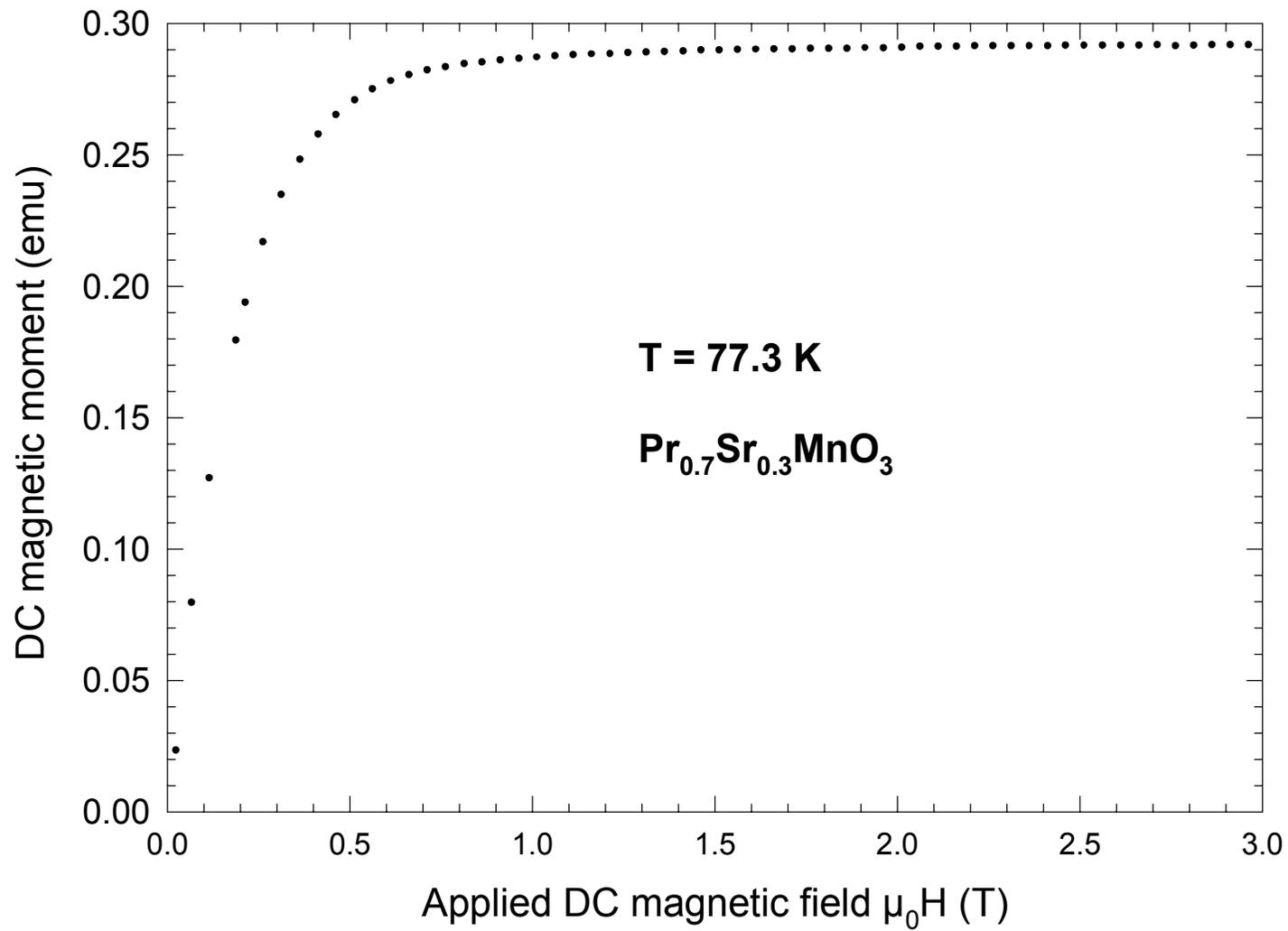
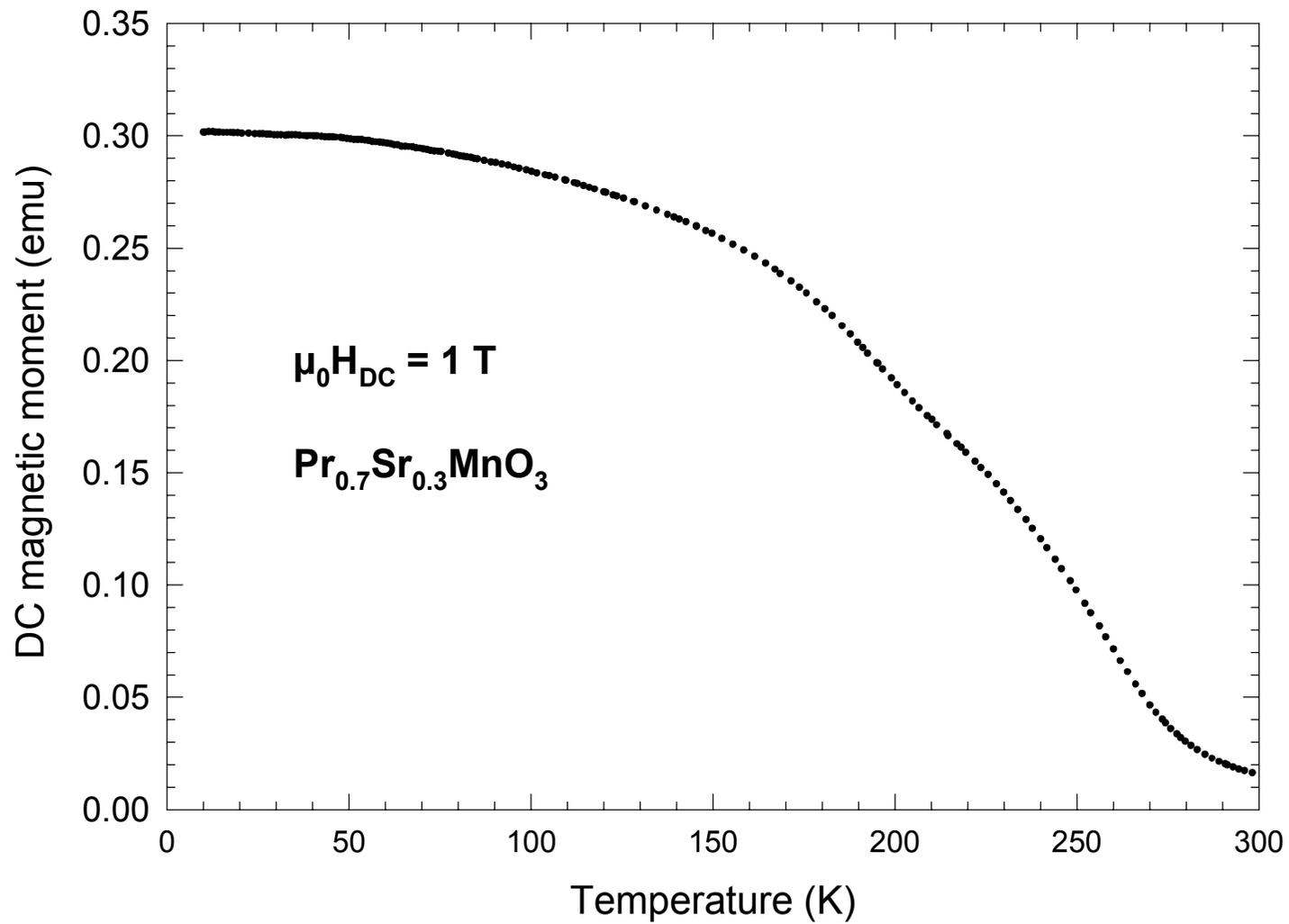
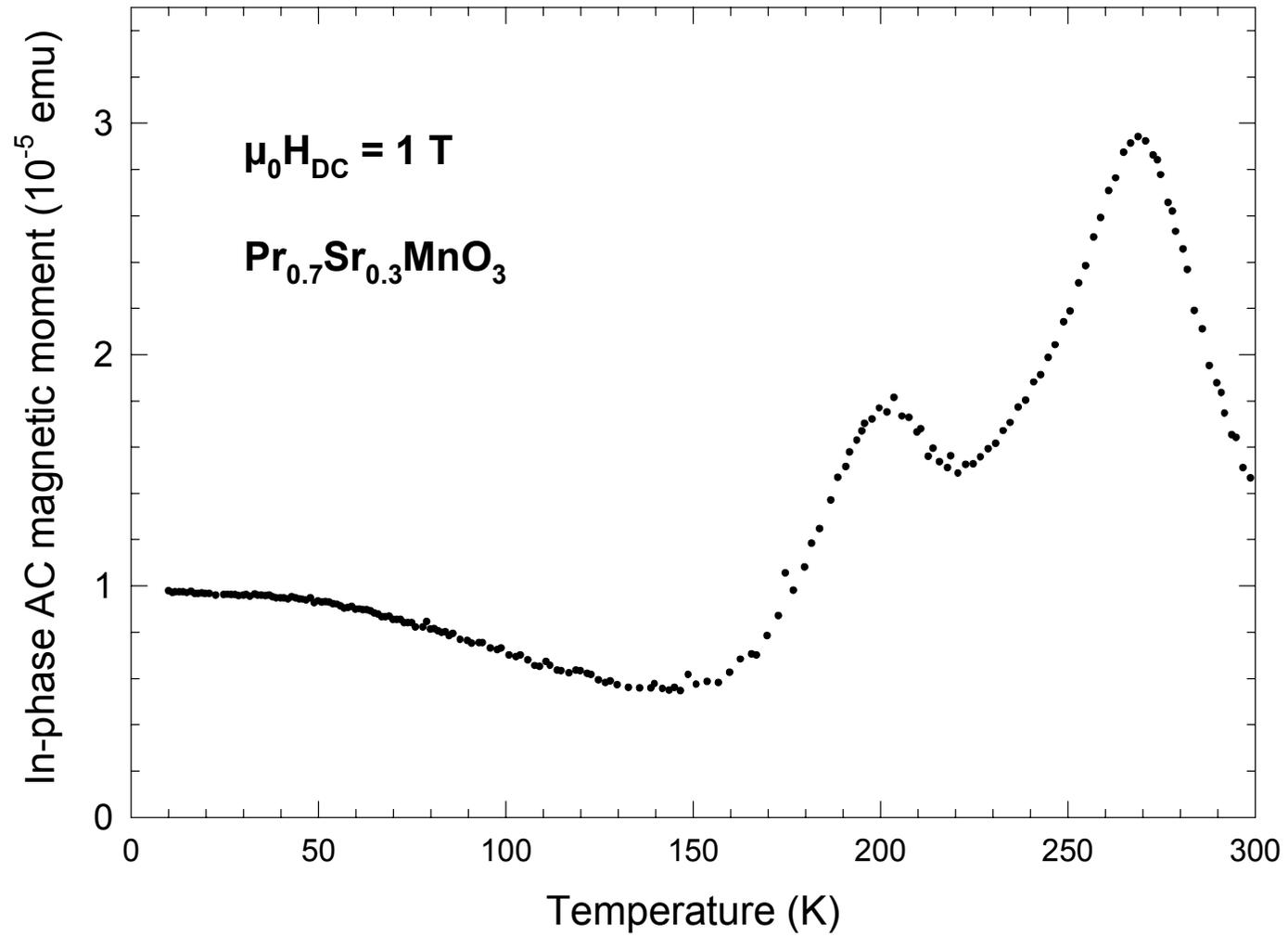
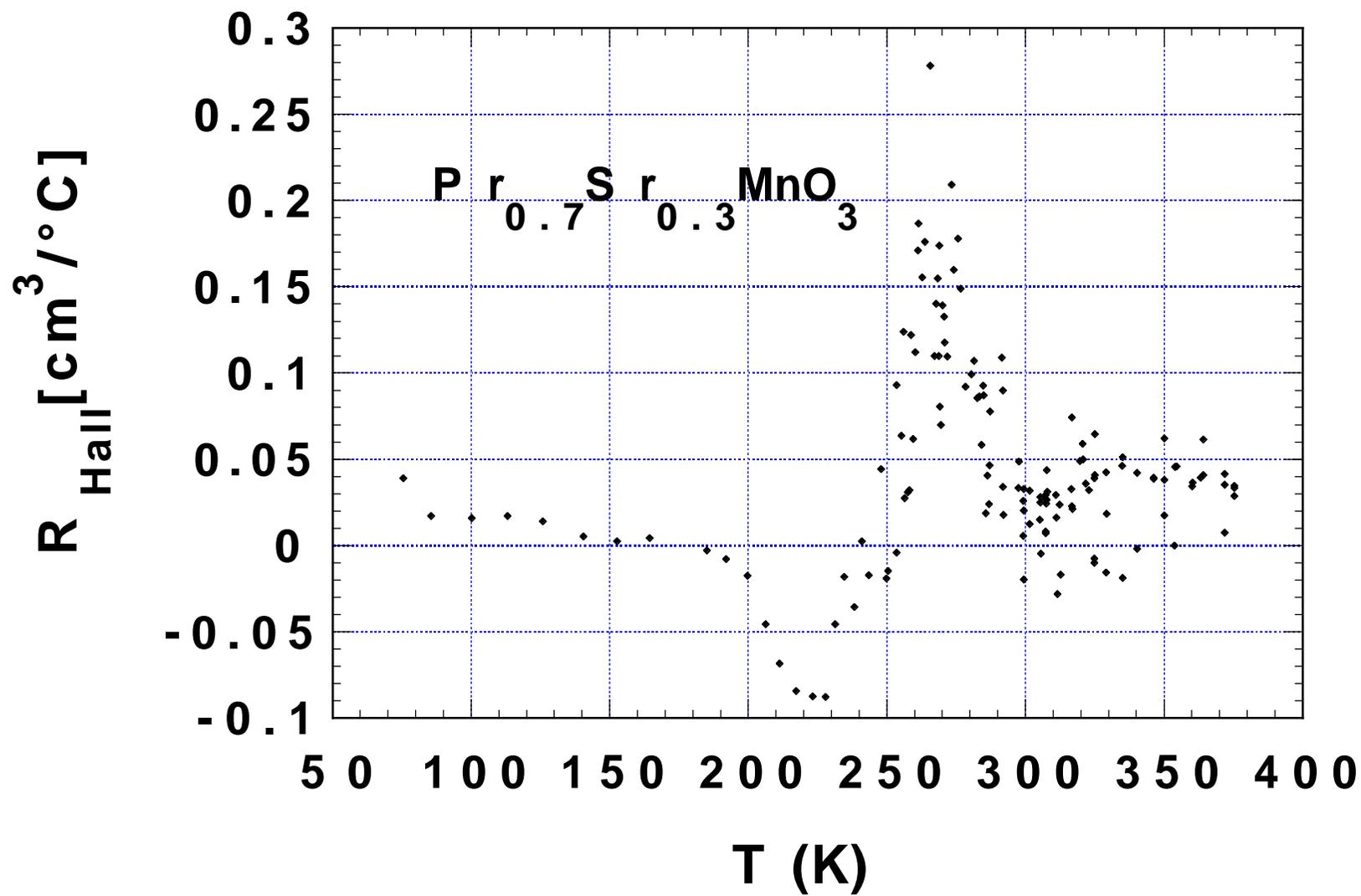
\end{document}